\pdfoutput=1

\documentclass[conference]{IEEEtran}
\IEEEoverridecommandlockouts
\usepackage{cite}
\usepackage{amsmath,amssymb,amsfonts}
\usepackage{graphicx}
\usepackage{multicol}
\usepackage{textcomp}
\usepackage[table]{xcolor}
\def\BibTeX{{\rm B\kern-.05em{\sc i\kern-.025em b}\kern-.08em
    T\kern-.1667em\lower.7ex\hbox{E}\kern-.125emX}}

\usepackage{makeidx}

\usepackage[utf8]{inputenc}
\usepackage{cmap}
\usepackage[T1]{fontenc}

\usepackage[strings]{underscore}
\usepackage{listings}
\usepackage[normalem]{ulem}
\usepackage{url}
\usepackage{color}

\usepackage[%
rm={oldstyle=false,proportional=true},%
sf={oldstyle=false,proportional=true},%
tt={oldstyle=false,proportional=true,variable=true},%
qt=false%
]{cfr-lm}

\usepackage{paralist}
\usepackage{csquotes}
\usepackage{microtype}

\makeatletter
\g@addto@macro{\UrlBreaks}{\UrlOrds}
\makeatother

\usepackage{pdfcomment}
\hypersetup{hidelinks,
   colorlinks=true,
   allcolors=black,
   pdfstartview=Fit,
   breaklinks=true
}
\usepackage[all]{hypcap}

\usepackage[capitalise,nameinlink]{cleveref}
\crefname{section}{Sect.}{Sect.}
\Crefname{section}{Section}{Sections}

\usepackage{xspace}
\newcommand{\eg}{e.\,g.,~}
\newcommand{\ie}{i.\,e.,~}

\definecolor{bluekeywords}{rgb}{0.13,0.13,1}
\definecolor{greencomments}{rgb}{0,0.5,0}
\definecolor{redstrings}{rgb}{0.9,0,0}

\definecolor{ltblue}{rgb}{0,0.4,0.4}
\definecolor{dkblue}{rgb}{0,0.1,0.6}
\definecolor{dkgreen}{rgb}{0,0.35,0}
\definecolor{dkviolet}{rgb}{0.3,0,0.5}
\definecolor{dkred}{rgb}{0.5,0,0}

\lstdefinelanguage{Coq}{ 
mathescape=true,
texcl=false, 
morekeywords=[1]{Section, Module, End, Require, Import, Export,
  Variable, Variables, Parameter, Parameters, Axiom, Hypothesis,
  Hypotheses, Notation, Local, Tactic, Reserved, Scope, Open, Close,
  Bind, Delimit, Definition, Let, Ltac, Fixpoint, CoFixpoint, Add,
  Morphism, Relation, Implicit, Arguments, Unset, Contextual,
  Strict, Prenex, Implicits, Inductive, CoInductive, Record,
  Structure, Canonical, Coercion, Context, Class, Global, Instance,
  Program, Infix, Theorem, Lemma, Corollary, Proposition, Fact,
  Remark, Example, Proof, Goal, Save, Qed, Defined, Hint, Resolve,
  Rewrite, View, Search, Show, Print, Printing, All, Eval, Check,
  Projections, inside, outside, Def},
morekeywords=[2]{forall, exists, exists2, fun, fix, cofix, struct,
  match, with, end, as, in, return, let, if, is, then, else, for, of,
  nosimpl, when},
morekeywords=[3]{Type, Prop, Set, true, false, option},
morekeywords=[4]{pose, set, move, case, elim, apply, clear, hnf,
  intro, intros, generalize, rename, pattern, after, destruct,
  induction, using, refine, inversion, injection, rewrite, congr,
  unlock, compute, ring, field, fourier, replace, fold, unfold,
  change, cutrewrite, simpl, have, suff, wlog, suffices, without,
  loss, nat_norm, assert, cut, trivial, revert, bool_congr, nat_congr,
  symmetry, transitivity, auto, split, left, right, autorewrite},
morekeywords=[5]{by, done, exact, reflexivity, tauto, romega, omega,
  assumption, solve, contradiction, discriminate},
morekeywords=[6]{do, last, first, try, idtac, repeat},
morecomment=[s]{(*}{*)},
showstringspaces=false,
morestring=[b]",
morestring=[d],
tabsize=3,
extendedchars=false,
sensitive=true,
breaklines=false,
basicstyle=\small,
captionpos=b,
columns=[l]flexible,
identifierstyle={\ttfamily\color{black}},
keywordstyle=[1]{\ttfamily\color{dkviolet}},
keywordstyle=[2]{\ttfamily\color{dkgreen}},
keywordstyle=[3]{\ttfamily\color{ltblue}},
keywordstyle=[4]{\ttfamily\color{dkblue}},
keywordstyle=[5]{\ttfamily\color{dkred}},
stringstyle=\ttfamily,
commentstyle={\ttfamily\color{dkgreen}},
literate=
    {forall}{{\color{dkgreen}{$\forall\;$}}}1
    {exists}{{$\exists\;$}}1
    {<-}{{$\leftarrow\;$}}1
    {<=}{{$\le$}}1
    {>=}{{$\ge$}}1
    {<>}{{$\neq$}}1
    {=>}{{$\Rightarrow\;$}}1
    {==}{{\code{==}\;}}1
    {==>}{{\code{==>}\;}}1
    {->}{{$\rightarrow\;$}}1
    {<->}{{$\leftrightarrow\;$}}1
    {<==}{{$\leq\;$}}1
    {\#}{{$^\star$}}1 
    {\\o}{{$\circ\;$}}1 
    {\@}{{$\cdot$}}1 
    {\/\\}{{$\wedge\;$}}1
    {\\\/}{{$\vee\;$}}1
    {++}{{\code{++}}}1
    {...}{{$\ldots$}}1
    {~}{{$\neg\;$}}1
    {\@\@}{{$@$}}1
    {\\mapsto}{{$\mapsto\;$}}1
    {\\hline}{{\rule{\linewidth}{0.5pt}}}1
}[keywords,comments,strings]

\usepackage{enumitem}
\setlist[itemize]{noitemsep, topsep=0pt}
\setlist[enumerate]{noitemsep, topsep=0pt}

\let\llncssubparagraph\subparagraph
\let\subparagraph\paragraph
\usepackage{titlesec}
\let\subparagraph\llncssubparagraph

\titlespacing{\section}{0pt}{10pt plus 0pt minus 2pt}{8pt plus 0pt minus 2pt}
\titlespacing{\subsection}{0pt}{8pt plus 0pt minus 1pt}{6pt plus 0pt minus 1pt}

\begin{document}


\author{\IEEEauthorblockN{Grigoriy Volkov}
\IEEEauthorblockA{\textit{Faculty of Computer Science} \\
\textit{National Research University}\\
\textit{Higher School of Economics}\\
Moscow, Russia \\
gdvolkov@edu.hse.ru}
\and
\IEEEauthorblockN{Mikhail Mandrykin}
\IEEEauthorblockA{\textit{Software Engineering Department} \\
\textit{Ivannikov Institute for System Programming of the}\\
\textit{Russian Academy of Sciences}\\
Moscow, Russia \\
mandrykin@ispras.ru}
\and
\IEEEauthorblockN{Denis Efremov}
\IEEEauthorblockA{\textit{Faculty of Computer Science} \\
\textit{National Research University}\\
\textit{Higher School of Economics}\\
Moscow, Russia \\
defremov@hse.ru}
}

\setlength{\abovedisplayskip}{1pt plus 1pt}
\setlength{\belowdisplayskip}{1pt plus 1pt}

\title{Lemma Functions for Frama-C:\\ C Programs as Proofs}

\maketitle

\begin{abstract}
   This paper describes the development of an auto-active verification technique in the Frama-C framework. We outline the lemma functions method and present the corresponding ACSL extension, its implementation in Frama-C, and evaluation on a set of string-manipulating functions from the Linux kernel. We illustrate the benefits our approach can bring concerning the effort required to prove lemmas, compared to the approach based on interactive provers such as Coq. Current limitations of the method and its implementation are discussed.
\end{abstract}


\setcounter{footnote}{0}

\section{Introduction}\label{sec:introduction}
Deductive verification is one of the most ``heavyweight'' static techniques of formal reasoning about software properties. It requires manual annotation of source code with formal specifications and manual or semi-automatic proof of conformance between code and specifications. Deductive verification is applied in areas where the cost of error can be too high. The main benefit of this approach is that it can mathematically guarantee that code is free from particular error classes.

In the toolsets based on the Hoare logic (such as Frama-C~\cite{framac}, VeriFast~\cite{verifast2011},  Dafny~\cite{leino2010dafny}, GNATprove~\cite{carre1990spark}),
in order to verify the correctness of a function, the verification engineer is required to define a precondition --- a predicate that should hold in order for the following code to be functionally correct and contain no runtime errors (\ie models of undefined behavior that are known to the verification tool), a postcondition --- a predicate that should hold after a function execution and describe what this code should act like to be functionally correct. Pre- and postconditions form the function's contract. Frame-conditions or effects are sometimes viewed as a separate entity in the contract, but typically they present a type of postcondition that restrict the memory state of a program. A formal contract is required to reason about the function's correctness.

Given a formal contract, an engineer can try to prove that a function conforms to its specification. To do this one needs to additionally specify loop invariants and termination expression (loop variants) in case the function under analysis contains loops. Then the verification instrument generates verification conditions (VCs), which can be separated into categories: safety and behavioral ones. Safety VCs are responsible for runtime error checks for known (modeled) types, while behavioral VCs represent the functional correctness checks. All VCs need to be discharged in order for the function to be considered fully proved (totally correct). This can be achieved either by manually proving 
all VCs discharged with an interactive prover (\ie Coq, Isabelle/HOL or PVS) or with the help of automatic provers.

It is well known that most problems related to verifying the conformance between some code in basically any practical programming language and its formal specification in any sufficiently expressive specification language are undecidable in general. The situation when an automatic prover cannot discharge a verification condition is common. This may be due to various reasons: limited time or memory resources for the prover, an error in the code, an incorrect formal specification, an incomplete formal specification, an incapacity of the prover to finish the proof. The specifications debugging is hard. Despite the modern advances~\cite{z3-counter-example,Petiot2018,Hentschel2018} in the field, the whole situation is still poor when specifications become complex.

As it can be seen the whole process of deductive verification requires a significant manual effort from a verification engineer. This implies that practically all verification tools have some capabilities for user interaction beyond simply supplying the program and its specification. This additional user interaction required to eventually prove the correspondence between the code and its specification is usually called the proof overhead. This overhead can significantly depend on the way the user interaction is accomplished in a particular verification tool. Even though these ways can generally be very diverse, for simplicity we dare to speculatively group them into the following three major categories:
\begin{itemize}
    \item External provers. The vast majority of verification tools support this approach to user interaction at some point, be it as a part of a standard workflow or as a last resort. The essence is that the verification conditions (VCs) produced by the tool are translated into the native language of some interactive proof assistant, \eg Coq, Isabelle/HOL or PVS. Then the resulting theorems are proved using the capabilities of the prover, sometimes with some extensions provided by the verification tool, \eg custom tactics. This approach is often both simple and efficient, but for some use cases, it has significant drawbacks such as the ones we describe in this paper.
    \item Interactive proof editors. Some verification tools such as WP plugin for Frama-C, Why3 verification platform or KIV~\cite{kiv} verification system go further and implement their own custom interactive proof management facilities such as tactics, transformations, and strategies. This is a hard yet powerful approach with its own advantages and shortcomings. One of its major disadvantages especially relevant to our particular case is a high implementation overhead, which prevents tentative adoption and quick preliminary evaluation of the results.
    \item Auto-active proofs. One of the possible techniques to reduce and automate the manual work and make deductive verification more suitable for such methods as continuous verification~\cite{moving-fast-with-sf,cv-amazon-s2n,cv-facebook-infer} is auto-active verification. The general idea is to express the proofs in the same input language as the program and its specification themselves relying on the Curry-Howard correspondence between the programs and the proofs. This approach is quite widely applied in particular in such verification tools as Dafny~\cite{leino2010dafny}, VeriFast~\cite{verifast2011}, GNATprove~\cite{adaredblack}, and Why3~\cite{why3ide}. This is the approach that we pursue and evaluate in this paper.
\end{itemize}

In this paper, we present our experience with integration of support for auto-active verification in ACSL~\cite{acsllang}, Frama-C, and AstraVer deductive verification tool and applying this approach to our existing set of specified string-manipulating functions from the Linux kernel. We demonstrate that integration of auto-active proofs to ACSL and Frama-C is both beneficial and promising approach to automated deductive verification that significantly reduces manual proof overhead.

This paper is organized as follows. 
In~\Cref{sec:verker} we give an overview of the VerKer project and its specifications and describe the problems we are solving by applying auto-active verification to this project. Next, in~\Cref{sec:auto-active-proofs} we present the implementation of the technique in the Frama-C framework and the AstraVer plugin. In~\Cref{sec:results,sec:open-issues} we describe the obtained result and discuss future work on open issues.

\section{VerKer Project}\label{sec:verker}
The VerKer project~\cite{isolaverker} consists of standard library functions taken from the Linux kernel. The main idea of the project is to develop the correct and detailed ACSL~\cite{acsllang} specifications and fully prove these functions with the Frama-C framework. Thus, evaluating the toolset (Frama-C + AstraVer Plugin (a fork of Jessie) + Why3 + Solvers) on the real code (not artificial examples) and highlighting C idioms that are hard to describe with ACSL specification, not possible to model with the deductive verification plugin, or just giving the right direction for the tools development.

ACSL is a Behavioral Interface Specification Language (BISL)~\cite{bisl} implemented in Frama-C. At the same time, many higher-level formalisms for efficient specification and verification of heap-manipulating or parallel programs such as dynamic frames, separation logic, permissions, rely/guarantee reasoning (\eg ownership methodology), specification refinement (\eg function contract inclusion) or auto-active verification are not included in ACSL. Therefore ACSL is not tightly bound to any particular verification methodology and support for its features can significantly vary in tools.


\subsection{Verification Approach}\label{sec:approach}
Specifications development is an iterative process. It cannot be broadly described in the general case. Every algorithm requires its own approach~\cite{two-lines-of-code-with-why3,why3-binary-heaps,ghosts-for-lists2018,acslbyexample} with separate axiomatizations fitting each particular case and implementation. A common way to write formal specifications is to factor out the most complex logical statements into theorems and lemmas so that the provers can discharge the rest of the verification conditions automatically. The theorems and lemmas are then proved manually with an interactive prover. The main benefits of this approach are the following:
\begin{itemize}
    \item Lemmas are formulated in a more general way so that they contain less implementation context specific details (\eg the assertion in a particular function). This eases the process of their manual proving and allows one to reuse them in multiple proofs of VCs and other lemmas.
    \item In this case, lemmas are ``aside from code'', thus making formal specifications more maintainable and tolerable to small code changes because they rarely provoke changes in general statements.
\end{itemize}
The VerKer project adheres to this approach in the ACSL specifications. The project currently contains 26 fully proved functions from Linux kernel library folder, 16 functions among them are string related (\ie have the str* prefix and perform operations on strings), and 7 are memory related (\ie have mem* prefix). Only 9 functions (see \Cref{table:verker-lemmas}) contain in its formal specification a logical function with lemmas formulated on it. The formal contracts are written in a way to prove correspondence between logical functions and C functions. The lemmas are excessive, and provers sometimes can automatically prove the correctness of a lemma based on other lemmas.


In the paper~\cite{isolaverker} it is claimed that the functions are fully-proved but lemmas are not proved (only automatic checks for contradiction was performed). The proving of all lemmas in the project considered as a future work in the paper.

\begin{table}
    \begin{center}
        \begin{tabular}{| l | c | c | c |}
            \hline
                \multicolumn{1}{|p{2cm}|}{\centering Function \\ Name } & \multicolumn{1}{|p{1.4cm}|}{\centering Number of \\ Lemmas } & \multicolumn{1}{|p{2cm}|}{\centering Proved \\ Automatically } & Unproved \\
                \hline
                \hline
                check_bytes8 & 3  & 3  & 0 \\
                strlen       & 10 & 4  & 6 \\
                skip_spaces  & 7  & 1  & 6 \\
                strchr       & 7  & 4  & 3 \\
                strchrnul    & 7  & 5  & 2 \\
                strspn       & 8  & 5  & 3 \\
                strcspn      & 5  & 2  & 3 \\
                strnlen      & 17 & 11 & 6 \\
                strpbrk      & 5  & 2  & 3 \\
                \hline
                \hline
                Total        & 69 & 37 & 32 \\
                \hline
        \end{tabular}
    \end{center}
    \caption{Lemmas in the VerKer project}
    \label{table:verker-lemmas}
\end{table}

\subsection{Shortcomings and limitations}
Thus we can found the following limitations of the approach based on external provers:
\begin{itemize}
  \item The use of external interactive proof assistants or proof editors requires the users to get accustomed to at least two different specification representations, corresponding languages and tools. This makes it harder for the users to employ their skills acquired during the process of specification within one verification environment (\eg the deductive verification plugin) later in the process of proof within the other one (\eg the Coq proof assistant). In particular, the range of supported language constructs can significantly differ. Many ACSL constructs such as pointer dereference do not have direct counterparts in the language of the interactive prover.
  \item The reproduction of external proofs requires the use of external tools that work on their own representations --- yet the translations performed by the verification tool in order to obtain the external representations can cause various surprising instabilities, where the proof suddenly fails after small supposedly irrelevant changes are made to program source code or specifications. This is especially well demonstrated by the pervasive problem of generating unique, stable and predictable names for various program and specification entities such as variables and lemmas. For instance, when a lemma or variable is renamed, the proofs have to be adjusted accordingly, yet the relation between the names used in the original program and the names used in the proofs can be quite complex.
  \item The external proofs are kept separately from the specifications and are therefore not directly available to the reader of the verified program. This can make it harder to estimate the effort needed for modification of certain parts of the program or specification or to keep the proofs in sync with the program and specification.
  \item The use of external interactive provers prevents an application of various encodings optimized for automatic provers, \eg SMT-solvers since the VCs should remain manageable by different tools that apply different tactics or decision procedures. In particular, while SMT solvers efficiently handle large ground (quantifier-free) formulas modulo theories, most interactive tools are much more suitable to handle smaller formulas in higher-order logic.
  \item External proofs introduce additional problems of proof location and sharing when the same lemma is included or imported in several source code files.
\end{itemize}

\section{Auto-active Proofs}\label{sec:auto-active-proofs}
 According to its original definition in~\cite{autoactive}, \emph{auto-active verification} is any verification technique where user input is supplied before VC generation. The user input can be given in various forms including but not limited to: auxiliary annotations such as verifier pragmas, triggers, ad-hoc lemma instantiations, ghost code, and lemma functions. Among those, lemma functions are a primary tool intended to convey proofs of user-defined auxiliary lemmas. The proofs are expressed as ordinary pure (\ie without side-effects) total (\ie always successfully terminating) imperative functions. Though the execution of pure total imperative code does not produce any observable effects with respect to the programming language semantics, when interpreted by the verification tool the instructions of the code can produce a substantial effect on its internal state or the internal state of the automatic theorem prover. In particular, through the process of VCs generation, the proof-carrying code can determine the fragment of current proof context that is explicitly known to the verifier or the prover versus the fragment that is only known implicitly as a set of derivable inferences. As most of the logic relevant to deductive verification is generally undecidable, neither the verifier nor the prover can ever explore all propositions derivable in the current context. The code of the lemma functions thus serves as a guide for the verification tool that efficiently indicates the relevant parts of the search space. One of the most significant cases where verification tools need such a guide is with inductive proofs as they have very limited support in automatic provers.
 
 Consider the following inductive proof expressed as a lemma function:
 \[
   \begin{array}{l}
    \textbf{logic char} *\textit{strchrnul}(\textbf{char} *s, \textbf{char} ~ c) =\\
    \quad *s \equiv c \mathrel{?} s :\\
    \quad *s \equiv {'}\backslash{0}{'} \mathrel{?} s :\\
    \quad \quad strchrnul(s + 1, c);\\
    \ldots\\
   \end{array}
 \]
 \[
    \begin{array}{l}
    \text{/*@ \textbf{ghost}}\\
    \text{~~~@ /@ \textbf{lemma}}\\
    \text{~~~@  ~@ \textbf{requires}}~valid\_str(s);\\
    \text{~~~@  ~@ \textbf{decreases}}~strlen(s);\\
    \text{~~~@  ~@ \textbf{ensures}}  ~s \le strchrnul(s, c) \le str + strlen(s);\\
    \text{~~~@  ~@/}\\
    \text{~~~@ \textbf{void}}~strchrnul\_in\_range(\textbf{char}*s, \textbf{char}~c)\\
    \text{~~~@ \{}\\
    \text{~~~@   ~~\textbf{if}}~(*s \mathrel{!=} {'}\backslash{0}{'} \mathrel{\&\&} *s \mathrel{!=} c)\\
    \text{~~~@}     ~~~~strchrnul\_in\_range(s + 1, c);\\
    \text{~~~@ \}}\\
    \text{~~~@*/}
  \end{array}
 \]
The lemma and its proof are expressed within a single recursive function definition $strchrnul\_in\_range$. The induction here is simple as it exactly follows the recursive definition of the logic function $strchnul$. The pre-condition of the function requires the parameter string $s$ to be a valid C string so that it has a well-defined finite length. Then the \textbf{if} statement in the function body splits between the base and inductive cases. In the base cases $*s \mathrel{==} {'}\backslash{0}{'}$ and $*s \mathrel{==} c$ the post-condition trivially follows from the definition of the function $strchrnul$. In the inductive case, which corresponds to the body of the \textbf{if} statement, the post-condition follows from the inductive assumption for strings of a smaller length and an additional instance of a lemma about the length of the string $s + 1$ that is implicitly present in the proof context (this lemma is actually also proved by a lemma function that is provided earlier in the code). The recursion is well-founded since in the only recursive call the value of the variant $strlen(s)$ specified in the \textbf{decreases} clause is strictly smaller than that in the caller function.  With the AstraVer plugin, the correctness of the function $strchrnul\_in\_range$ can be verified automatically by discharging all the VCs using an SMT solver such as Alt-Ergo or CVC4. The automatic proofs require no induction and only a few quantifier instantiations, while all the necessary terms are explicitly present in the VCs. So in either solver, the verification takes far less than a second.

Now since the function $strchrnul\_in\_range$ is proved correct for arbitrary values of its parameters, and since the function is pure and total, the statement about its correctness with respect to the contract can be expressed as the following lemma:
\[
  \begin{array}{l}
  \textbf{lemma}~strchrnul\_in\_range:\\
   \quad \forall \textbf{char}*s, \textbf{char}~c;\\
   \quad \quad valid\_str(s) \implies\\
   \quad \quad \quad s \le strchrnul(s, c) \le str + strlen(s);
  \end{array}
\]
This lemma can be automatically generated from the function contract and used in the proofs later in the program code, both explicitly by the user by calling the function $strchrnul\_in\_range$ with some concrete arguments or implicitly by the SMT solver by instantiating the lemma with some concrete terms.

This is the basic idea of lemma functions and their use. Now let us consider our approach to the integration of lemma function into ACSL in particular. We first present our existing approach and current results that we obtained by using it, and then we discuss some of its limitations and possible directions of future developments.

\subsection{Integrating Lemma Functions Into ACSL/Frama-C}

In our fork of the Frama-C platform developed as part of the AstraVer toolset, we implemented relatively simple support for lemma functions as a special case of total pure ghost functions. Both ACSL specification and its implementation in mainline Frama-C provide support for ghost functions. Meanwhile, the     AstraVer plugin is able to generate VCs to check ACSL \textbf{assigns}, \textbf{allocates}, \textbf{decreases} and \textbf{terminates} clauses of function contracts. So the only missing part to provide basic support for lemma functions was the automatic generation of lemmas and their implicit importing in all the code functions located further in the source code. 

More concretely, in order to allow lemma functions to be used in a similar way to regular ACSL lemmas (which can only be written in logic), we introduced a new \texttt{lemma} keyword for C function specifications.

Our modified version of Frama-C processes it in the following way:
\begin{itemize}
    \item For a function contract of the form
    \[
       \begin{array}{l}
       \text{/@}~\textbf{requires}~R(p_1, \ldots, p_n, g_1, \ldots, g_n);\\
       \text{~@}~\textbf{ensures}~E(p_1, \ldots, p_n, g_1, \ldots, g_n, \backslash{result});\\
       \text{~@/},
       \end{array}
    \]
    where $p_1, \ldots, p_n$ are function parameters, $g_1, \ldots, g_n$ are global variables and the special variable $\backslash{result}$ denotes the return value of the function,
    a corresponding axiom is generated, that is, an axiom of the form
    \[
       \begin{array}{l}
         \forall \texttt{typeof(}p_1\texttt{)}~p_1, \ldots, \texttt{typeof(}p_n\texttt{)}~p_n,\\
         \quad \texttt{typeof(}g_1\texttt{)}~g_1, \ldots, \texttt{typeof(}g_n\texttt{)}~g_n;\\
         \quad \quad  \exists \texttt{\texttt{typeof(}}\backslash{result}\texttt{)}~ \backslash{result};\\
         \quad \quad \quad R(p_1, \ldots, p_n, g_1, \ldots, g_n) \implies {}\\
         \quad \quad \quad \quad E(p_1, \ldots, p_n, g_1, \ldots, g_n, \backslash{result});
       \end{array}
    \]
    \item All logic definitions such as functions, predicates, axioms, and lemmas in ACSL are grouped into axiomatic blocks. The generated axiom is placed in a new axiomatic block. AstraVer implements on-demand import of all the definitions from axiomatic blocks based on occurrences of symbols defined in the axiomatic. As the axiom name cannot be directly used in an ACSL specification, the axiom is not included in any proof context until some other symbol in this new axiomatic is defined and later used in another specification (code function or another axiomatic block).  In particular, this approach removes the axiom from the context of the lemma function itself, preventing the axiom from trivially proving the function's verification conditions.
    \item A dummy identically true predicate with a unique name is generated, and its definition is added into the new axiomatic block. 
    \item A precondition requiring the dummy predicate from the new axiomatic block is inserted into the contracts of all function located after the currently processed lemma function, which causes the definitions from the axiomatic block (including the generated axiom) to be imported into the proof context of these functions. As the dummy predicate is generated by the tool and has a unique name, the user cannot access it and thus create any extra dependencies disrupting the ordering of lemma proofs according to their code locations.
    \item The lemma function itself gets some additional clauses inserted into its contract specification:
    \begin{itemize}
        \item $\textbf{assigns}~\backslash{nothing};~~\textbf{allocates}~\backslash{nothing}$~--- to ensure the function is pure (\ie it does not touch any global state);
        \item $\textbf{terminates}~\backslash{true};$~-- to prevent abrupt termination (\ie calls to \texttt{exit()}, \texttt{abort()} etc.)~--- loop and recursion termination is already required by default in the AstraVer plugin.
    \end{itemize}
\end{itemize}

An alternative approach enforcing the topological ordering of lemma functions has been considered: explicit removal of the axiom from the context of the functions located before the lemma body.
This is possible to achieve using the \texttt{meta remove_prop} pragma in the generated Why3ML model. However, this would have required Frama-C to track the connection between the function and the axiom explicitly, and the AstraVer plugin to use that connection to generate the \texttt{meta} statement in the WhyML output. A minimal implementation that fully translates lemma functions into existing specification entities was chosen in the end.

\section{Results}\label{sec:results}

\begin{figure*}
    \centering
    \begin{tabular}{cc}
    \begin{tabular}{l}
    \begin{lstlisting}[language=Coq,basicstyle=\scriptsize]
Theorem Strchr_skipped :
  forall voidP_str_0_30_alloc_table_at_L: alloc_table voidP,
  forall charP_charM_str_0_30_at_L : map (pointer voidP) Int8.t,
  forall str_0_0 : pointer voidP,
  forall c_1_0 : Int8.t,
  forall i_0_0 : Uint64.t,
    valid_str
      str_0_0
      voidP_str_0_30_alloc_table_at_L
      charP_charM_str_0_30_at_L /\
    strchr
      str_0_0
      c_1_0
      charP_charM_str_0_30_at_L <> (null : pointer voidP) /\
    Uint64.infix_lseq (Uint64.of_int 0%Z) i_0_0 /\
    (Uint64.to_int i_0_0 < sub_pointer
                             (strchr                           $\smash{\left.
           \vphantom{\begin{array}{l}A\\[33em]A\end{array}}
         \right\}\rotatebox[origin=c]{90}{\normalsize Generated by Why3}}$
                                str_0_0
                                c_1_0
                                charP_charM_str_0_30_at_L)
                              str_0_0))%Z ->
    get
      charP_charM_str_0_30_at_L
      (shift str_0_0 (Uint64.to_int i_0_0)) <> c_1_0.
  (* Why3 intros
    voidP_str_0_30_alloc_table_at_L
    charP_charM_str_0_30_at_L
    str_0_0 c_1_0 i_0_0
    (h1,(h2,(h3,(h4,h5)))). *)
  intros
    voidP_str_4_82_alloc_table_at_L
    charP_charM_str_4_82_at_L
    str_4 c_4_0 i_1
    (h1,(h2,(h3,(h4,h5)))).

  Definition P (i : int) :=
    forall a : alloc_table voidP,
    forall m : map (pointer voidP) Int8.t,
    forall s : pointer voidP,
    forall c : Int8.t,
      valid_str s a m /\
      strchr s c m <> (null : pointer  voidP) /\
      (0 <= i)%Z /\
      (i < sub_pointer (strchr s c m) s)%Z ->
      get m (shift s i) <> c.
  specialize Wf_Z.natlike_rec3 with (P := P) as Ind.    
  assert (P 0) as Base.
    unfold P.
    intros a m s c Valid.
    unfold not.
    \end{lstlisting}\\
    $\ldots \hspace{18em} \smash{\left.
           \vphantom{\vrule height 3.5em}
         \right\}\rotatebox[origin=c]{90}{{\normalsize 13} lines}}$\\
    {\begin{lstlisting}[firstnumber=58,language=Coq,basicstyle=\scriptsize]
    rewrite First in Valid.
    rewrite Sub_pointer_self in Valid.
    omega.                                                     $\smash{\left.
           \vphantom{\begin{array}{l}A\\[36.6em]A\end{array}}
         \right\}\rotatebox[origin=c]{90}{{\normalsize 93} lines of {\normalsize user proof}}}$
  assert (forall z : int, (0 < z)%Z -> P (Z.pred z) -> P z) as Step.
    intros z Gt0 Pred.
    unfold P.
    intros a m s c Valid.
    \end{lstlisting}}\\
    $\ldots \hspace{18em} \smash{\left.
           \vphantom{\vrule height 3.5em}
         \right\}\rotatebox[origin=c]{90}{{\normalsize 55} lines}}$\\
    {\begin{lstlisting}[firstnumber=114,language=Coq,basicstyle=\scriptsize]
    apply Strchr_same_block with
      (voidP_s_4_26_alloc_table_at_L := a).
    tauto.
  assert (forall z : int, (0 <= z)%Z -> P z) as Lemma.
    apply Ind. 1,2:tauto.
  unfold P in Lemma.
  apply Lemma with (a := voidP_str_4_82_alloc_table_at_L).
    unfold Uint64.infix_lseq in h3.
    rewrite Uint64.Of_int in h3.
    \end{lstlisting}}\\
    $\ldots\hspace{18em} \smash{\left.
           \vphantom{\vrule height 2.5em}
         \right\}\rotatebox[origin=c]{90}{{\normalsize 8} lines}}$\\
    {\begin{lstlisting}[firstnumber=127,language=Coq,basicstyle=\scriptsize]
    unfold Uint64.in_bounds. omega.
    tauto.
Qed.
    \end{lstlisting}}
    \end{tabular}
    &
    \hspace{3em}
    \begin{lstlisting}[language=C,basicstyle=\footnotesize,mathescape=true]
/$*$@ $\small\textbf{ghost}$
  @ /@ $\small\textbf{lemma}$
  @  @ $\small\textbf{requires}~~ valid\_str(str);$
  @  @ $\small\textbf{requires}~~ strchr(str, c) \neq \backslash{null};$
  @  @ $\small\textbf{requires}~~ 0 \le i < strchr(str, c) - str;$
  @  @ $\small\textbf{decreases}~ i;$
  @  @ $\small\textbf{ensures}~~~   str[i] \neq c;$
  @  @/
  @ void strchr_skipped(char *str, char c,
  @                     size_t i)
  @ {
  @   if (i > 0 && *str != '\0' && *str != c)
  @     strchr_skipped(str + 1, c, i - 1);
  @ }
  @*/
    \end{lstlisting}
    \end{tabular}
    \caption{Proofs of lemma \texttt{strchr_skipped}. Coq proof on the left. Auto-active proof with lemma function on the right.}
    \label{fig:coq-vs-lemma-function}
\end{figure*}
Even though at the previous stage of out project~\cite{isolaverker} we ended up with 32 unproved lemmas, we did not attempt to match our lemma function proofs at this stage exactly with the lemmas used at the previous stage. Instead, the goal was to ensure all the code can be verified without resorting to any additional assumptions by completely replacing all the lemmas involved with a minimum required set of lemma functions. Thus we ended up with 31 lemma functions required to fully prove the VCs for some code functions as presented in~\Cref{table:verker-lemma-functions}.

Unfortunately, a thorough comparison of the proof overhead between the interactive and auto-active proofs turned out to be very hard. The current support for interactive proofs that existed in AstraVer plugin was quite basic, which resulted in considerably large manual proofs. The reduction of these proofs seems only possible with very significant implementation effort including implementation of relevant custom tactics and optimizing the VC generation for manual proofs. For the reference, we provide an illustrative comparison of an interactive auto-active proof of a single lemma in~\Cref{fig:coq-vs-lemma-function} as it can be performed with the current implementation. This example shows the reduction in proof overhead of more than tenfold, yet a different more optimized manual proof augmented with appropriate custom tactics could potentially result in a much smaller number.

However, in the context of our project, the use of lemma functions became a decisive choice that enabled complete, practical, stable and future-proof verification of auxiliary lemmas with minimal implementation effort. The ability to instantiate lemmas by explicitly calling the corresponding lemma functions and the availability of all general C constructs in the proof code provided the verification engineer with sufficient control. By these means in our examples we were able to reduce the verification time of each lemma function (cumulative time of the solvers on all function's VCs) to a few seconds.
    \begin{table}
      \begin{center}
        \begin{tabular}{| l | c | c |}
            \hline
                \multicolumn{1}{|p{2cm}|}{\centering Function \\ Name } & 
                \multicolumn{1}{|p{2cm}|}{\centering Number of \\ Ghost Functions } &
                \multicolumn{1}{|p{2cm}|}{\centering Number of \\ Lemma Functions } \\
                \hline
                \hline
                strlen       & 4 & 2 \\
                skip_spaces  & 1 & 3 \\
                strchr       & 0 & 5 \\
                strchrnul    & 0 & 5 \\
                strspn       & 1 & 5 \\
                strcspn      & 0 & 3 \\
                strnlen      & 3 & 6 \\
                strpbrk      & 0 & 2 \\
                \hline
                \hline
                Total        & 9 & 31 \\
                \hline
        \end{tabular}
    \end{center}
    \caption{Lemma functions}
    \label{table:verker-lemma-functions}
\end{table}

\section{Open Issues and Future Work}\label{sec:open-issues}

\subsection{Logic types and expressions in code}
Unlike many specification languages originally developed with auto-active verification in mind, ACSL maintains a clear distinction between code and specification languages. In particular, C and logic types as well as C and logic expressions are represented entirely separately within the Frama-C kernel and are only allowed in their respective positions. Therefore lemma functions cannot be directly used to prove the lemmas with quantification over logic types. This limitation can be overcome in individual cases of quantified logic variables ranging over the domains bounded by C values.

For universal quantification of the form
\[\forall~\mathbb{L}~v;~ v_1 \le v < v_2 \implies P(v),\]
where $v_1$ and $v_2$ are C values of type $\mathbb{C}$ a proof can be accomplished through the use of an auxiliary lemma
\[\text{aux:~}\forall~\mathbb{C}~v;~ v_1 \le v < v_2 \implies P(v),\]
which can then be generalized using a dummy proof of the form
\[
  \begin{array}{l}
  \text{/@ \textbf{lemma}}\\
  \text{ @ \textbf{ensures} } \forall~\mathbb{L}~v;~ v_1 \le v < v_2 \implies P(v);\\
  \text{ @/}\\
  \text{\textbf{void} gen() \{}\\
  \text{~~/@ \textbf{loop invariant} } v_1 \le i \le v_2;\\
  \text{~~ @ \textbf{loop invariant} }
    \forall~\mathbb{C}~j;~ v_1 \le j < i \implies P(j);\\
  \text{~~ @/}\\
  \text{~~\textbf{for} (}\mathbb{C}~\text{i = v1; i < v2; i++)}\\
  \text{~~~~}aux(i);\\
  \text{\}}
  \end{array}
\]  

For logic expressions $l(\overline{u})$ ranging over the C values $v_1 \le l(\overline{u}) < v_2$, where $\overline{u}$ are C expressions, a proxy C function can be introduced:

\[
\begin{array}{l}
  \text{/*@ \textbf{ensures}}~\backslash{result} \equiv l(\overline{u});\\
  \text{~~ @ \textbf{assigns}}~\backslash{nothing};\\
  \text{~~ @*/}\\
  \mathbb{C}~proxy(\overline{u});
\end{array}
\]
Note that $l(\overline{u})$ may contain logic constructs that are thus encapsulated by the proxy. In general, proxies should be introduced carefully as their post-conditions are not checked and may introduce inconsistencies, but in some cases, the proxy can be given an implementation (definition) that satisfies the corresponding specification. In our proofs, we only introduced proxies with definitions.

Ultimately, the proper support for logic types and expressions in the ghost code should be added to ACSL specification and implemented in Frama-C to fully support auto-active proofs with lemma functions.
\subsection{Lemma functions and axiomatics}
Normally, lemmas in ACSL are grouped into axiomatic blocks. However, axiomatic blocks cannot include code. Therefore, lemma functions cannot be directly grouped into axiomatic blocks. Moreover, allowing lemma functions to be declared as lemmas in axiomatic blocks introduces a problem with ordering of proofs. To be well-founded, the lemma functions should be split into strongly connected components and topologically ordered. However, unlike code functions that can only be called explicitly, lemmas can be instantiated implicitly, provided they are preset in the proof context. When each lemma function has just a single definition, the definition order of lemma functions can serve as their required topological order, with an only additional restriction that mutually recursive functions should be somehow declared simultaneously. However, when lemma functions are allowed to have separate declarations and definitions, the ordering of lemma functions becomes generally unspecified. In the current implementation, the lemma functions are allowed to have separate declarations, but they are later grouped into strongly connected components, and the definition of the first function in the component is then used for topological ordering. This essentially simulates simultaneous definitions. However, this approach is completely ignorant of the lemma declarations grouped into axiomatic blocks. Ultimately, instead of choosing between the declaration and the definition points to be used for topological ordering, we propose including the lemma function definitions (bodies) into the axiomatic blocks directly. The ordering then can be retained as in the current implementation. However, this would be a significant change to the ACSL specification.
\subsection{Proof reuse and higher-order logic}
 Careful consideration of the resulting auto-active proofs from our study reveals that many of them are very similar to each other and represent instances of some more general proof patterns. In general, it is natural to express such more general lemmas and their proofs using higher-order logic (HOL). However, the support for higher-order reasoning in SMT-based deductive verification tools such as Jessie or AstraVer is usually very limited due to the inherent incompleteness of decision procedures for HOLs. An alternative solution to direct support for HOL would be a stratified solution, where the core specification language does not support any HOL constructs, but a separate higher-level language such as the language of modules does provide constructs that can be used to express and prove higher-order propositions. This approach is already used in Why3ML specification language.
 Let us illustrate this approach on two example lemmas from our study:
  \[
   \begin{array}{l}
    \text{\textbf{logic char}} *\textit{strchr}(\textbf{char} *s, \textbf{char } c) =\\
    \quad *s \equiv c \mathrel{?} s :\\
    \quad *s \equiv {'}\backslash{0}{'} \mathrel{?} (\text{char} *) \backslash{null} :
    strchr(s + 1, c);\\
    \textbf{lemma} ~ \textit{strchr\_skipped}:\\
    \quad \forall \textbf{char} *s, c, \text{size\_t}~i;\\
    \quad \quad \textit{valid\_str}(s) \land {}\\
    \quad \quad \textit{strchr}(s, c) \not\equiv \backslash{null} \land {}\\
    \quad \quad 0 \le i < \textit{strchr}(s, c) - s \implies\\
    \quad \quad s[i] \not\equiv c;\\
    \ldots\\
    \textbf{logic char} *\textit{strchrnul}(\textbf{char} *s, \textbf{char} ~ c) =\\
    \quad *s \equiv c \mathrel{?} s :\\
    \quad *s \equiv {'}\backslash{0}{'} \mathrel{?} s :
    strchrnul(s + 1, c);\\
    \textbf{lemma} ~ \textit{strchrnul\_skipped}:\\
    \quad \forall \textbf{char} *s, c, \text{size\_t}~i;\\
    \quad \quad \textit{valid\_str}(s) \land {}\\
    \quad \quad 0 \le i < \textit{strchrnul}(s, c) - s \implies\\
    \quad \quad s[i] \not\equiv c;
   \end{array}
 \]
Currently, the lemmas have to be proved as two separate lemma functions. Though there is a single generic function that suits both cases, its definition makes use of higher-order features :
\[
   \begin{array}{l}
    \begin{aligned}
    \textbf{logic char} *\textit{strchrcomb}(&\text{boolean} ~ cond(\textbf{char} *s, \textbf{char} ~ c),\\
    &\textbf{char} *dft(\textbf{char} *s),\\
    &\textbf{char} *s, \textbf{char} ~ c) =
    \end{aligned}\\
    \; cond(s, c) \mathrel{?} s :
    *s \equiv {'}\backslash{0}{'} \mathrel{?} dft(s, c) :
    strchrcomb(s + 1, c);
   \end{array}
\]
Within the stratified approach, this function should be placed in a separate module alongside the functions corresponding to its parameters, auxiliary definitions, and appropriate lemmas:
\[
   \begin{array}{l}
     \textbf{predicate} ~ cond(\textbf{char} *s, \textbf{char} ~ c)~\textbf{reads} *s;\\
     \textbf{logic char} *dft(\textbf{char} *s);\\
     \textbf{predicate} ~ dft\_same = \forall \textbf{char} *s; ~dft(s) \equiv s;\\
     \textbf{predicate} ~ dft\_null = \forall \textbf{char} *s; ~dft(s) \equiv \backslash{null};\\
     \textbf{logic char} *strchrcomb(\textbf{char} *s, \textbf{char} ~ c) =\\
     \quad cond(s, c) \mathrel{?} s :\\
     \quad *s \equiv {'}\backslash{0}{'} \mathrel{?} dft(s) :\\
     \quad \quad strchrcomb(s + 1, c);\\
     \textbf{lemma}~strchrcomb\_skipped:\\
     \quad \forall \textbf{char} *s, c,~\text{size\_t}~i;\\
     \quad \quad valid\_str(s) \land {}\\
     \quad \quad (dft\_same \lor dft\_null \land strchrcomb(s, c) \not\equiv \backslash{null}) \land {}\\
     \quad \quad 0 \le i < \textit{strchrcomb}(s, c) - s \implies {}\\
     \quad \quad !cond(s + i, c);
   \end{array}
\]
Then the lemmas can be proved in the general case as usual, \eg by using lemma-functions. The resulting module can be instantiated by providing concrete definitions for logic functions $cond$ and $dft$ to obtain both logic functions $strchr$ and $strchrnul$. Upon instantiation, their corresponding lemmas do not have to be proven, and they can be safely instantiated directly into axioms.



\section{Conclusion}\label{sec:conclusion}
In this paper we propose the implementation of auto-active proofs technique in the Frama-C framework and AstraVer plugin, evaluate our implementation on VerKer project specifications by proving all string-related lemmas, and demonstrate the benefits of the approach by comparing it with manual lemma proofs in Coq.

The results of the work are publicly available. The source code with specifications and verification protocols are available as a part of the VerKer project. The proofs are easy reproducible since there is a configuration of a continuous integration system. The modified toolset is available on the AstraVer project page.

The obtained results show the approach is very promising and should be adopted in the mainline of AstraVer toolset. Auto-active verification in Frama-C ease the work of a verification engineer, make specifications more maintainable and potentially can lower the cost of deductive verification.

\bibliographystyle{IEEEtran}
\bibliography{FRAMAC}
\end{document}